\documentclass[english,pra,preprint,a4paper,floats,showpacs]{revtex4}
\usepackage{graphicx}
\usepackage{amssymb}

\makeatletter

\usepackage{babel}
\makeatother
\begin{document}

\title{Dark Optical Lattice of Ring Traps for Cold Atoms}

\author{Emmanuel Courtade, Olivier Houde, Jean-François Cl\'{e}ment, Philippe
Verkerk and Daniel Hennequin}

\email{daniel.hennequin@univ-lille1.fr}

\affiliation{Laboratoire de Physique des Lasers, Atomes et Mol\'{e}cules, UMR CNRS,
Centre d'\'Etudes et de Recherche Lasers et Applications, Universit\'{e}
des Sciences et Technologies de Lille, F-59655 Villeneuve d'Ascq cedex,
France}

\date{\today}

\begin{abstract}
We propose a new geometry of optical lattice for cold atoms, namely
a lattice made of a 1D stack of dark ring traps. It is obtained through
the interference pattern of a standard Gaussian beam with a counter-propagating
hollow beam obtained using a setup with two conical lenses. The traps
of the resulting lattice are characterized by a high confinement and
a filling rate much larger than unity, even if loaded with cold atoms
from a MOT. We have implemented this system experimentally, and demonstrated
its feasibility. Applications in statistical physics, quantum computing
and Bose-Einstein condensate dynamics are conceivable
\end{abstract}

\pacs{32.80.Pj, 39.25.+k}

\maketitle
Optical lattices provide a versatile tool to study the dynamical properties
of cold and ultracold atoms. They are presently the topic of intense
research activity, in particular because they represent an outstanding
toy model for various domains. In statistical physics, cold atoms
in optical lattices, through their tunability, made possible the observation
of the transition between Gaussian and power-law tail distributions,
in particular the Tsallis distributions \cite{tsallis}. Condensed
matter systems and strongly correlated cold atoms in optical lattices
offer deep similarities, as in the superfluid-Mott insulator quantum
phase transition \cite{mott}, in the Tonks-Girardeau regime \cite{tonks}
or for the emergence of a macroscopic current in the periodic potentials
\cite{current}. In quantum computing, optical lattices appear to
be an efficient implementation of a Feynman's universal quantum simulator
\cite{toolbox}, and are among the most promising candidates for the
realization of a quantum computer \cite{qcomp1}.

One of the main advantages of the optical lattices is their high flexibility.
By varying the shape of the lattice, a wide range of configurations
is reached. Currently, many studies deal with 1D lattices, in particular
because quantum effects are stronger in low-dimensional systems \cite{1D1}.
A particularly interesting situation concerns 1D lattices with periodic
boundary conditions, because many new effects appear \cite{dring1}.
Recently, an experimental implementation has been proposed, where
the lattice sites are distributed along rings \cite{dring3}. In a
more complex configuration, the sites themselves could have the shape
of a ring, allowing e.g. the study of solitons in 1D Bose-Einstein
Condensates (BECs) with periodic boundary conditions \cite{1DBEC1}
or atomic-phase interferences between such BECs \cite{1DBEC2}. Experimental
realization of such 1D rings is still an open question, both as a
lattice or as a single trap. Large magnetic single ring traps have
been produced, in connection with the study of the atomic Sagnac effect
\cite{sring1,sring2}, but their transverse confinement is weak, and
they cannot be considered as 1D rings. A more promising proposition
is an optical trap built with twisted light obtained from two counter-propagating
Laguerre-Gaussian beams with an azimuthal phase dependence \cite{twisted}.
Authors suggest that an optical lattice of such ring traps could be
created by combining several twisted molasses. Such an arrangement
has the drawback to trap the atoms where the light intensity is maximum.
This may result in serious perturbations of the atoms due to the trapping
beams \cite{dark}. In particular, some applications as quantum computation
require to trap the atoms in dark lattices, to make the system robust
against decoherence \cite{graph}. Contrary to bright lattices, where
even a 1D configuration leads to 3D trapping, 1D dark lattices do
not trap atoms in 3D: only 3D dark lattices trap atoms in 3D \cite{dark}.
So far, the only proposal for such a lattice consists in a gaussian
beam making a round trip in a confocal cavity \cite{cavity}. A difference
of waist between the two directions of propagation lead to a lattice
of ring traps with a $\lambda/2$ periodicity, where $\lambda$ is
the optical wavelength. Such a device has the advantage to generate
high light intensity inside the cavity, and so deeper traps than with
free-propagating beams. But this is obtained at the cost of the flexibility:
for example, changing the ring radius requires to change the cavity
mirrors. These difficulties probably explain why, to our knowledge,
this proposal has not yet been realized experimentaly.

We propose here a new geometry \textbf{\large }for a dark lattice
of ring traps, obtained from a hollow beam and a counter-propagating
gaussian beam, without any cavity. The radius and the thickness of
the rings can be adjusted indepently, and due to the stiff edges of
the hollow beam, the trap steepness is much larger than in \cite{cavity}.
Finally, the filling rate of each site is much larger than unity,
even when loaded with a Magneto-Optical Trap (MOT), contrary to 3D
dark lattices which require the use of a BEC. The filling rate should
even reach values in excess of 1000 atoms per site if an adequate
sequence is used to turn on the lattice.

The paper is organized as follows: we first discuss the principle
of the lattice of ring traps, then describe the experimental realization,
and finally show preliminary results concerning cold atoms loaded
in the lattice.

\begin{figure}
\includegraphics[width=1\columnwidth,keepaspectratio]{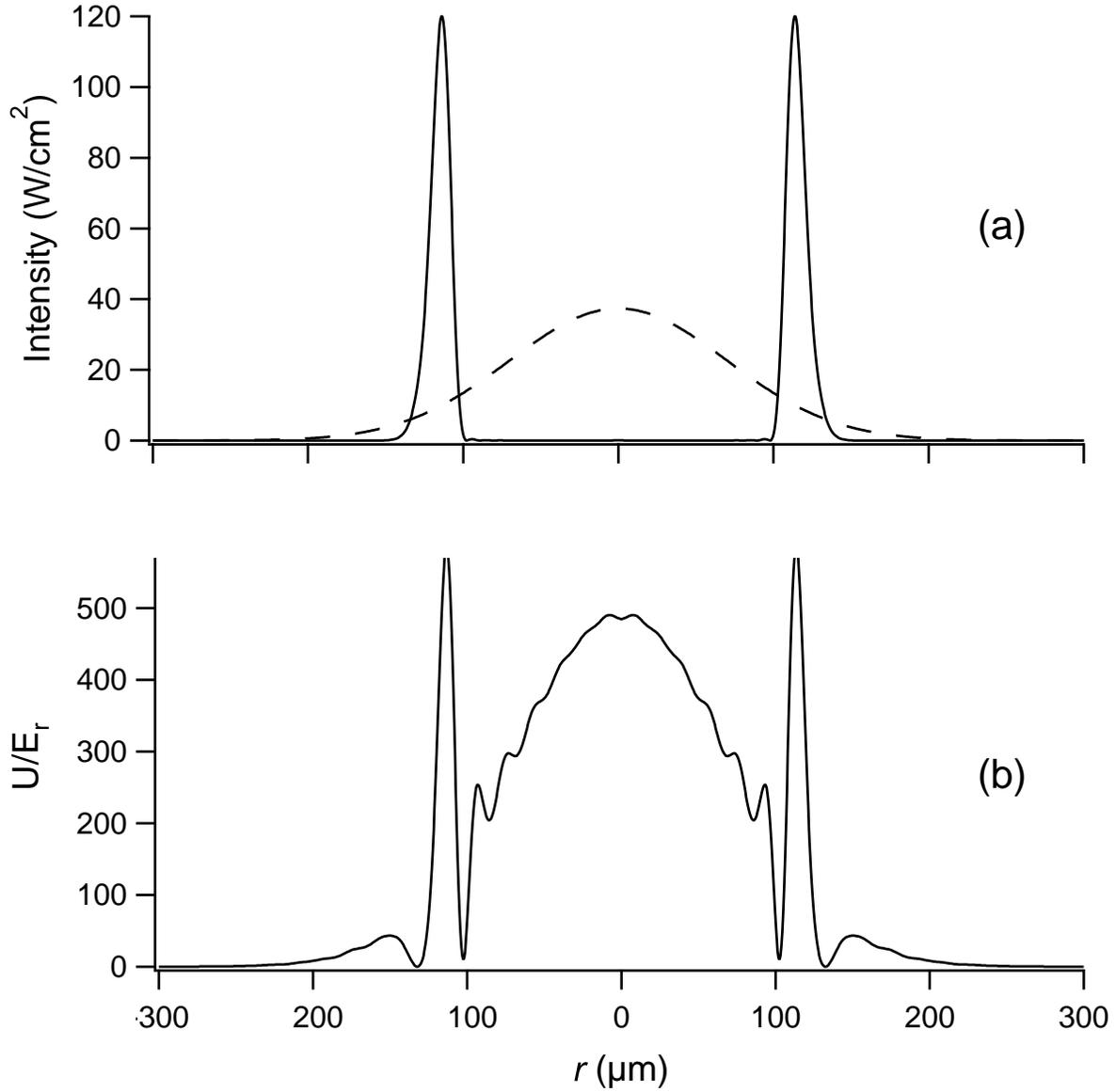}

\caption{\label{fig:pot}In (a), theoretical transverse profile of the gaussian
(dashed) and hollow (full) beams used in the experiment. In (b), the
resulting transverse potential at the bottom of the well. Parameters
are those used in the experiments. }
\end{figure}

Each individual trap is a 3D dark ring, and the lattice is a 1D stack
of such rings. Thus, the global shape of the potential is a bright
full cylinder with a pile of ring wells inside. To obtain this potential,
a standard gaussian beam interferes with a counterpropagating hollow
beam with no azimuthal phase dependence \cite{axicon}, contrary to
Laguerre-Gaussian beams used e.g. as waveguides \cite{waveguide}.
Both beams have the same blue detuned frequency, so that the trapping
sites correspond to the zero intensity places. Both beams propagate
along the $z$ vertical axis, and the hollow beam is a cylindrical
beam, with an intensity distribution along the radial direction $r$
as illustrated in Fig. \ref{fig:pot}a (solid line). When the two
beams are out of phase, two pairs of zeros of intensity appear symmetrically
on each side of the center, in $r=100$\,\ensuremath{µ}m and $r=130$\,\ensuremath{µ}m
on Fig. \ref{fig:pot}b, where the two beam intensities are equal.
Because of the cylindrical symmetry, these zeros correspond to two
concentric rings along the azimuthal direction. On the contrary, when
the two beams have the same phase, the intensity profile reaches its
maximum. This interference pattern results in a potential $U$ which
is, in the limit of weak saturation and large detuning, proportional
to the light intensity $I$:\begin{equation}
U=\frac{\hbar}{8}\frac{\Gamma^{2}}{\Delta}\frac{I}{I_{S}}\end{equation}
where $\Delta$ is the detuning, $I_{S}$ the saturation intensity
and $\Gamma$ the width of the atomic transition. Because the outer
ring is shallow, only the inner ring, in $r=100$\,\ensuremath{µ}m,
is a trap.

A hollow beam as described above is easily produced by a conical lens
\cite{axicon}. Conical lenses are extensively used to produce Bessel-Gauss
beams \cite{besselgauss} or annular beams \cite{axicon78,annular}.
To generate an annular hollow beam, we use a converging lens $L$
to shape the incident gaussian beam, and then a conical lens. Each
incident ray is deviated towards the optical axis $z$ by the conical
lens, and thus the incident gaussian beam is transformed in a ring
\cite{axicon}. A second conical lens is used to collimate the radius
$r$ of the ring, so that after this second lens, $r$ becomes constant
with $z$. The resulting hollow beam has a radius $r$ which depends
only on the distance between the two conical lenses, while its thickness
$\Delta r$ depends on the focal length of $L$. Thus $r$ and $\Delta r$
are adjustable independently.

\begin{figure}
\includegraphics[width=1\columnwidth]{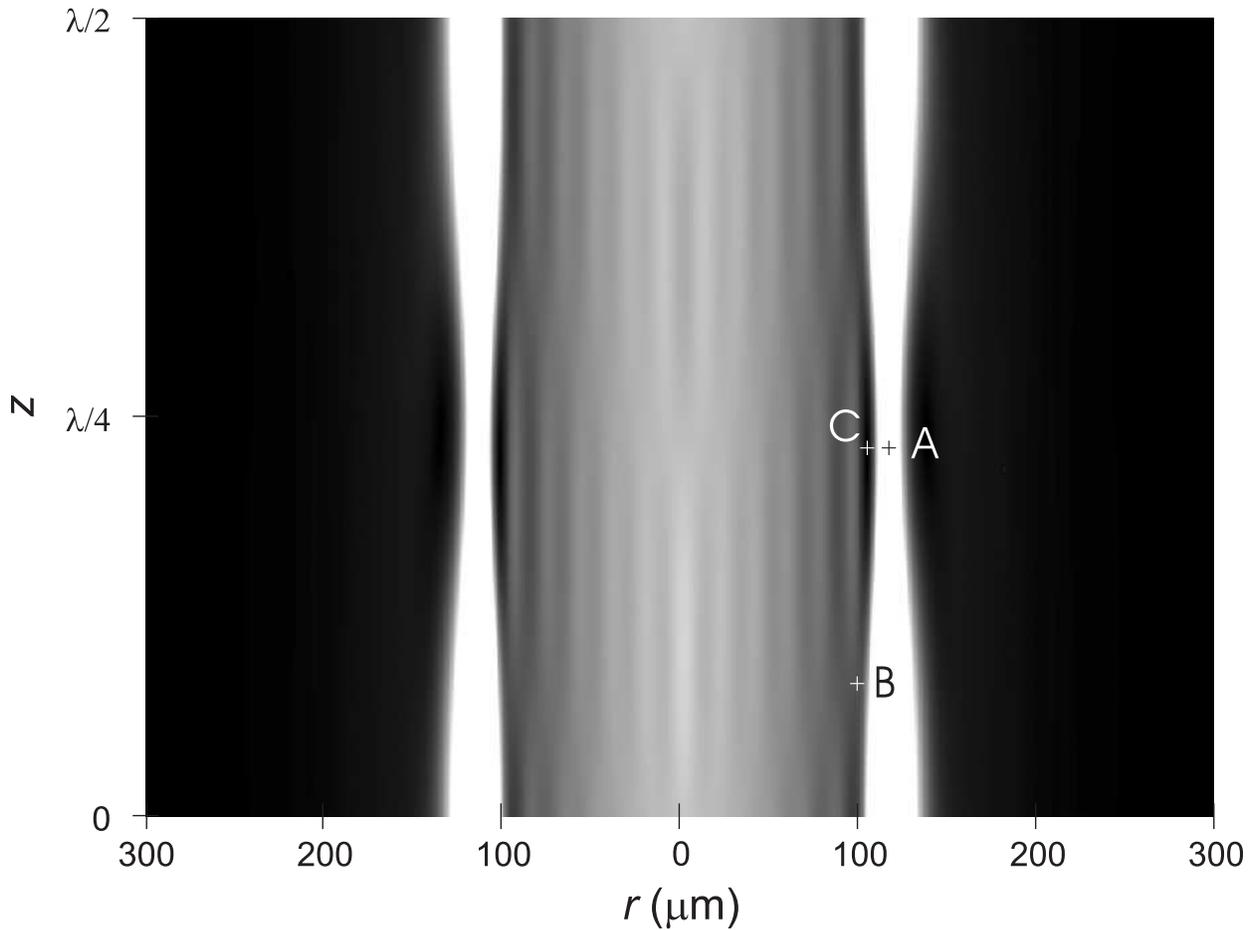}

\caption{\label{fig:pot2D}2D representation of the potential as a function
of the radius $r$ and the longitudinal coordinate $z$. The complete
potential has a revolution symmetry around the axis $r=0$. Parameters
are those of Fig. \protect \ref{fig:pot}. Note that the scales along
$z$ and $r$ are different. Dark corresponds to a zero potential.
Significance of points A, B and C is given in the text.}
\end{figure}

The potential is obtained by focusing the gaussian beam and the counterpropagating
hollow beam at the same point, so that the wave surfaces are planes
perpendicular to the propagation axis. The resulting potential has
a periodicity of $\lambda/2$, with a shape depending locally on the
phase $\phi$ between the two beams. It is illustrated through the
theoretical plots of Figs. \ref{fig:pot}b and \ref{fig:pot2D} where,
for sake of simplicity, we used the parameters of the experimental
demonstration described below: the hollow and gaussian intensities
are respectively $I_{H}=14\,\textnormal{mW}$ and $I_{G}=11.5\,\textrm{mW}$
(Fig. \ref{fig:pot}a), with $\Delta/2\pi=70\,\textrm{GHz}$. Fig.
\ref{fig:pot}b shows the potential transverse profile at the bottom
of the wells. The geometry of the ring appears clearly, with a confinement
of the order of $r/10$ for $U<200\, E_{r}$, and a height for the
external barrier of $580\, E_{r}$, where $E_{r}$ is the recoil energy.
Secondary minima, originating in the residual diffraction produced
by the mask used to remove inner rings of the hollow beam \cite{axicon},
appear inside the main ring, but because of their weak depth, they
should not be annoying in most applications.

A better understanding of the potential distribution can be obtained
from Fig. \ref{fig:pot2D}, where $U$ is plotted in gray scale versus
$r$ and $z$. The potential is periodic along $z$, with a period
$\lambda/2$. Atoms with low enough energy are confined in a torus
with a half-ellipse cross-section with axes of the order of 0.1 $\mu$m
and 10 $\mu$m, corresponding on the figure to the dark zone on the
point C. The height of the external barrier varies with $z$. The
minimum height $U_{A}=580\, E_{r}$, in point A of Fig. \ref{fig:pot2D},
occurs at the same $z$ as the bottom of the main well (Fig. \ref{fig:pot}b).
The internal barrier has a channel structure, with the lowest pass
in point B (Fig. \ref{fig:pot2D}), at a height of $U_{B}=200\, E_{r}$,
between two successive longitudinal sites.

The number of atoms that we should be able to put in each site of
this lattice depends of course on the density of the cloud of cold
atoms used to load the lattice, but also on the spatial overlap between
the cloud and the lattice. In particular, when the atoms are loaded
from a MOT, the capture volume of the lattice is decisive for its
filling rate. In the present case, the capture volume is determined
by the hollow beam diameter, which may be chosen of several hundreds
of $\mu$m. For example, with $r=100$ $\mu$m (Fig. \ref{fig:pot2D})
and an initial cloud of radius 1 mm, 1.5 \% of the initial atoms are
inside the hollow beam. Thus if the lattice is loaded with a cloud
of $10^{8}$ atoms in 4 mm$^{3}$, which are the typical characteristics
obtained from a MOT, $1.5\times10^{6}$ atoms are loaded in 2000 sites,
leading to a filling rate much larger than 1.

To test the feasibility of this lattice, we have implemented an experiment
with the characteristics described above. Cesium atoms are initially
cooled in a standard MOT with a $-3\Gamma$ detuning from resonance.
At time $t=-40$ ms, the magnetic field is turned off, while at time
$t=-30$ ms, the detuning is increased to $-5\Gamma$ and the trap
beam intensity is decreased: this sequence allows us to obtain at
time $t=0$ a 40 $\mu$K molasse, corresponding to an energy of $200\, E_{r}$,
with 10$^{8}$ atoms in typically 4~mm$^{3}$.

The hollow and gaussian beams are produced by two laser diodes injected
by a single master laser diode in an extended cavity, which ensure
the same frequency for both beams. For this demonstration, the beams
are tuned $70$ GHz above the atomic transition. In these conditions,
the power of the gaussian and hollow beams, which are respectively
11.5 and 14 mW, are sufficient to reach the needed potential depth
of $200\, E_{r}$. The gaussian beam has a minimum waist of 140 $\mu$m,
located at the level of the MOT. The axicon setup is mounted on an
optical rail, to guarantee a good stability of the beam. The incident
beam is collimated with a waist equal to 645 $\mu$m. The two conical
lenses, with a vertex angle of $2^{\circ}$, are separated by a distance
of about 10 cm, adjusted to obtain $r=1$ mm. The $L$ focal length
of 500 mm leads to $\Delta r=100$ $\mu$m. A telescope located just
before the trap reduces these values to $r\simeq100$ $\mu$m and
$\Delta r\simeq10$ $\mu$m. We obtain in the MOT a transverse distribution
of the hollow beam which is in excellent agreement with the theoretical
one.

\begin{figure}
\includegraphics[width=1\columnwidth,keepaspectratio]{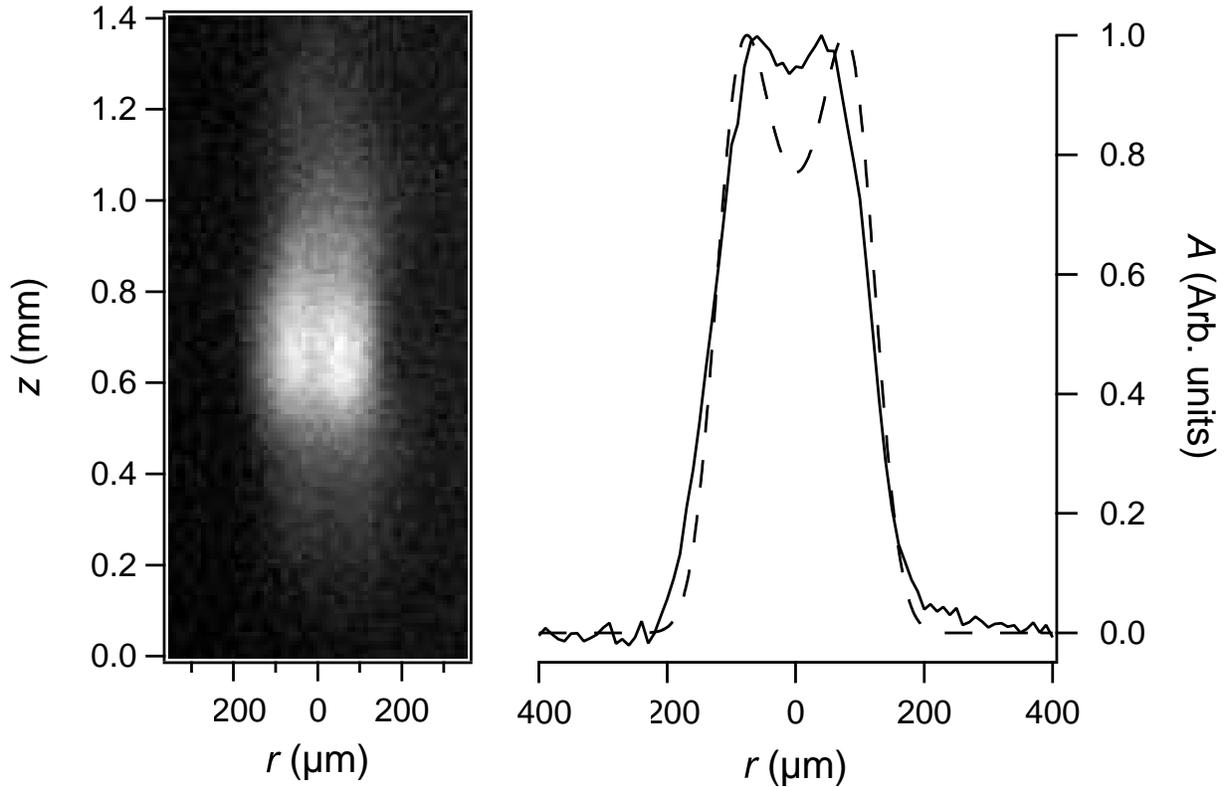}

\caption{\label{fig:picture}Left, side view of the lattice obtained by taking
a snapshot of the fluorescence of the atoms in the lattice at time
$t=40$~ms. Right, the corresponding transverse distribution (solid
line) and a rough estimate of the theoretical distribution (dashed
line). The two-bump structure reveals the annular structure of the
traps.}
\end{figure}

To load the cold atoms inside the lattice, the later is turned on
at a time $t<0$, so that when the molasse is switched off, the atoms
are already distributed inside the wells. At time $t=0$, the atoms
start to fall under the effect of gravity, except for those which
are trapped in the lattice. The free atoms need typically $25$ ms
to quit the camera field of view, so that for $t>25$ ms, only atoms
interacting with the optical lattice remain. To observe the atoms,
we switch on during 1 ms the trap laser beams near resonance, and
we used a low-noise cooled CCD camera to detect the fluorescence emitted
by the atoms. A typical picture is shown in Fig. \ref{fig:picture}.
As the resolution of the imaging setup is about 10 \ensuremath{µ}m,
individual rings separated by $\lambda/2$ cannot be distinguished.
On the contrary, the picture resolves the transverse distribution
caracterized by two maxima resulting from the side view of the rings.
The dashed line is the theoretical distribution obtained when the
potential is approximated to a double gaussian curve. Thus the experimental
distribution is less contrasted because the inner walls of the actual
wells are flatter than the outer ones.

\begin{figure}[t]
\includegraphics[width=1\columnwidth,keepaspectratio]{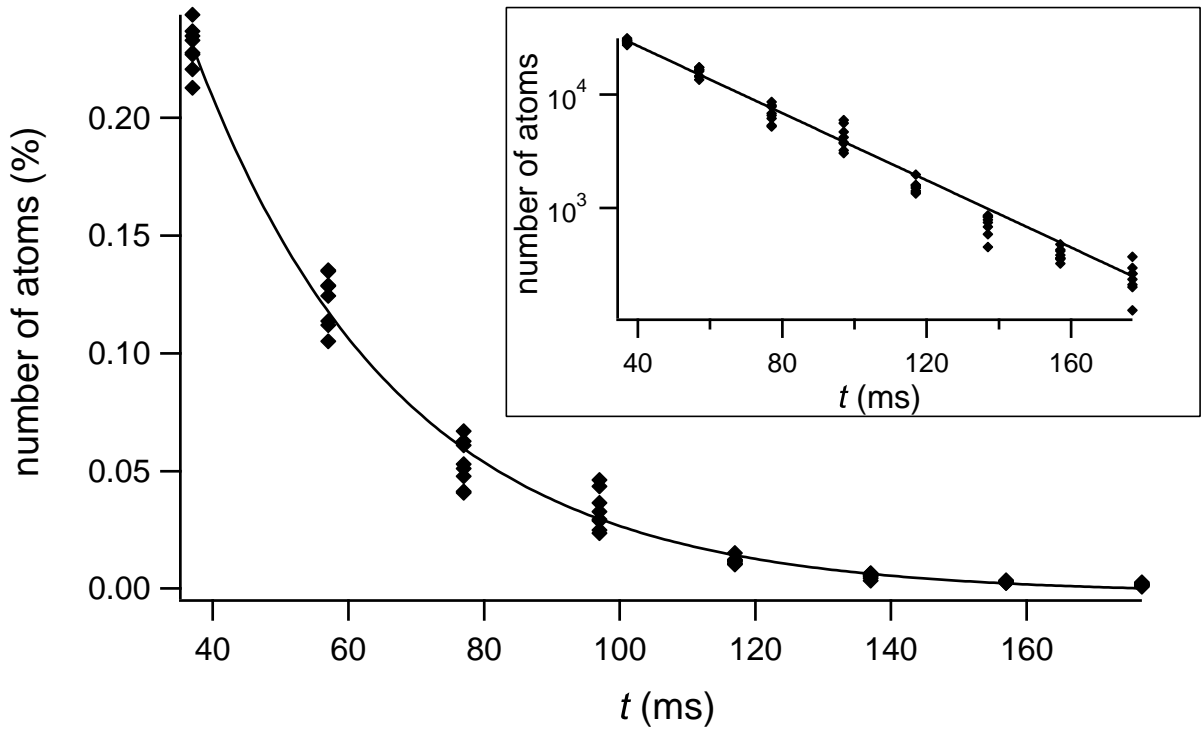}

\caption{\label{fig:life}Number of atoms in the lattice versus time. The
main plot is in linear scale, and the number of atoms is given as
a percentage of the molasse population. The solid line correspond
to the fit by an exponential with a decay time $\tau=30$ ms. In the
insert, the same results are shown in log scale, with the absolute
number of atoms. In both cases, points are experimental.}
\end{figure}

Fig. \ref{fig:life} shows the evolution of the population of the
lattice as a function of the time, for the above parameters. The points
are obtained by integrating the experimental pictures along the $z$
axis. In order to test the robustness of the procedure, eight measures
have been done for each point in abscissa. Fig. \ref{fig:life} shows
that the number of atoms decreases exponentially with time. The fit
on an exponential (solid line in Fig. \ref{fig:life}) gives a lifetime
of 30 ms. This value is in good agreement with the theoretical lifetime
of the atoms resulting from collisions and spontaneous emission. The
number of atoms in the lattice is also in good agreement with the
theoretical one: after 40 ms, there are still $30000$ atoms. Assuming
that these atoms are localized in about 700 lattice sites, we reach
a filling rate of 40 atoms per site. Note that actually, we have no
direct proof of the localization of the atoms, but only the periodic
optical potential can refrain the atoms from falling. However, only
a direct observation of atom localization, through e.g. Bragg diffraction,
would demonstrate unambiguously that atoms are trapped in the lattice.

In conclusion, we propose here a new lattice geometry, namely a 1D
stack of ring traps, and show its experimental feasibility. The experimental
set-up remains relatively simple, because the lattice is created from
only one pair of beams and it does not need any cavity, contrary to
the propositions respectively in \cite{twisted} and \cite{cavity}.
The other characteristics of this lattice are: high confinement of
the atoms due to the stiff walls of the trapping sites; large capture
volume and filling rate, due to the independence of the torus radius
and thickness; and weak interaction between light and atoms, as the
traps are dark. We implemented experimentally such a lattice, and
loaded it directly from a MOT. We measure a lifetime of the atoms
in the lattice of 30 ms. Although we have no direct evidence of the
localization of the atoms in the lattice sites, this demonstrates
the feasibility of this system. The lifetime should be improved by
changing some parameters which were not optimized for this feasibility
demonstration. For example, a decrease of the initial temperature
of the atoms in the molasse by an adequate cooling sequence lead to
relatively deeper traps. An increase of the detuning $\Delta$ of
the lattice, which requires more intense laser sources, reduces the
spontaneous emission. Finally, a decrease of the pressure of the thermal
atoms, through e.g. the use of a double cell, improves the collision
rate. Each of these enhancements will contribute to lengthen the lifetime
of the atoms in the lattice. If better filling rates are necessary,
it would also be possible to increase it. Indeed, when the lattice
is switched on, many atoms are heated on the lattice axis, because
this axis corresponds to a maximum of the potential. To avoid these
extra losses, we plan to switch on the lattice in two steps: first,
the hollow beam is switched on just after the trap beams are switched
off, so that the atoms inside the beam are trapped and remain in the
cylinder. Then, the gaussian beam is switched on progressively, so
that the atoms are adiabatically pushed in the ring traps. This precaution
prevents the atoms to be heated by a sudden increase of the potential. 

Finally, it would be interesting to study the dynamics of the atoms
in this novel geometry of lattice. Moreover, this lattice could be
used in systems where interactions between atoms in a same site or
in neighboring sites are required, or when periodic limit conditions
are necessary. 

The Laboratoire de Physique des Lasers, Atomes et Mol\'{e}cules is
Unit\'{e} Mixte de Recherche de l'Universit\'{e} de Lille 1 et du
CNRS (UMR 8523). The Centre d'\'Etudes et de Recherches Lasers et
Applications (CERLA) is supported by the Minist\`{e}re charg\'{e}
de la Recherche, the R\'{e}gion Nord-Pas de Calais and the Fonds
Europ\'{e}en de D\'{e}veloppement \'Economique des R\'{e}gions.
The group is supported by the Institut de Recherche sur les Composants
logiciels et mat\'{e}riels pour l'Information et la Communication Avanc\'{e}e
(IRCICA).

\end{document}